\documentclass[twocolumn,preprintnumbers,amsmath,amssymb,floatfix,prd]{revtex4}
\usepackage{epsfig}
\usepackage{dcolumn}
\usepackage{tabularx}
\usepackage{multirow}
\usepackage{hyperref}
\usepackage[dvipsnames]{xcolor}
\begin{document}
\title{Towards a Global QCD Analysis of Fragmentation Functions at Next-To-Next-To-Leading Order Accuracy}
\author{Ignacio Borsa}
\email{iborsa@df.uba.ar} 
\author{Rodolfo Sassot}
\email{sassot@df.uba.ar} 
\affiliation{Departamento de F\'{\i}sica and IFIBA,  Facultad de Ciencias Exactas y Naturales, Universidad de Buenos Aires, Ciudad Universitaria, Pabell\'on\ 1 (1428) Buenos Aires, Argentina}

\author{Daniel de Florian}
\email{deflo@unsam.edu.ar} 
\affiliation{International Center for Advanced Studies (ICAS) and IFICI, UNSAM, 
Campus Miguelete, 25 de Mayo y Francia (1650) Buenos Aires, Argentina}

\author{Marco Stratmann}
\email{marco.stratmann@uni-tuebingen.de}
\author{Werner Vogelsang}
\email{werner.vogelsang@uni-tuebingen.de}
\affiliation{Institute for Theoretical Physics, University of T\"ubingen, Auf der Morgenstelle 
14, 72076 T\"ubingen, Germany}

\begin{abstract}
We carry out a global QCD analysis of parton-to-pion fragmentation functions at next-to-next-to-leading order 
(NNLO) accuracy by performing a fit to the combined set of single-inclusive
electron-positron annihilation and, for the first time, semi-inclusive deep-inelastic scattering (SIDIS) multiplicity data.
For the latter, we utilize the approximate NNLO QCD corrections 
that were derived recently within the threshold resummation formalism. We explore the impact of the NNLO
corrections on the description of the SIDIS data sets in various kinematic regimes and on the resulting pion fragmentation functions.
\end{abstract}
%
\maketitle

{\it Introduction.---} 
Fragmentation functions (FF) 
constitute a crucial building block in 
perturbative calculations of scattering cross sections 
with detected final-state hadrons \cite{Collins:1981uw}.
In the presence of a sufficiently 
large energy scale, 
QCD factorization allows one 
to isolate the physics describing the transition 
of an outgoing parton to an observed, colorless hadron 
from the hard scattering process that produced the parton \cite{Collins:1989gx}.
The parton-to-hadron FFs, which precisely describe this transition are
a unique manifestation of the non-perturbative formation of QCD final states via ``hadronization'', and 
hence offer insights central to our knowledge and understanding of the theory of the strong interactions. 
The wide range of applications of FFs includes modern studies of the nucleon's spin structure in terms of polarized parton 
densities \cite{DeFlorian:2019xxt}, as well as investigations of modifications of hadron production
rates in scattering processes involving heavy nuclei \cite{Sassot:2009sh}.

FFs are process-independent quantities and thus can be determined from data 
by means of a global QCD analysis \cite{deFlorian:2007aj}. The energy scale dependence of FFs
can actually be computed perturbatively as an expansion in the strong coupling
and is currently known up to next-to-next-to-leading order (NNLO)
accuracy \cite{Mitov:2006ic,Almasy:2011eq}. Given the importance of FFs, their global analysis
has enjoyed considerable theoretical interest recently, resulting
in various new sets for different kinds of identified hadrons 
\cite{deFlorian:2014xna,Sato:2016wqj,deFlorian:2017lwf,Bertone:2017tyb,Anderle:2017cgl,Moffat:2021dji,
Khalek:2021gxf,Abdolmaleki:2021yjf,Borsa:2021ran}. 
But, unlike the case of parton distribution functions (PDFs), global analyses of FFs 
have so far been essentially limited to the next-to-leading order (NLO) accuracy of QCD perturbation theory.
This is due to the lack of  
NNLO computations for the cross sections for some of the most relevant processes 
sensitive to FFs, i.e., SIDIS and single-inclusive hadron production in proton-proton ($pp$) collisions.
As a result, the few NNLO extractions of FFs \cite{Anderle:2015lqa,Abdolmaleki:2021yjf} are currently based solely on
single-inclusive electron-positron annihilation (SIA) data. These studies found the NNLO
corrections to be small in the kinematic regime where SIA data exist. Ref. \cite{Anderle:2016czy}
extended the analysis of SIA by including all-order resummation effects at small hadron momentum fractions,
again observing very modest effects. 

The main purpose of our paper is to perform a first ``proof-of-principle'' {\it global} analysis of FFs at NNLO accuracy, 
based on the available SIA \cite{ref:siadata} and -- for the first time in an NNLO framework -- the SIDIS data for identified 
charged pions of \cite{HERMES:2012uyd,COMPASS:2016xvm}. Such an analysis has now become possible, thanks
to the recent derivation of approximate NNLO corrections for SIDIS in \cite{Abele:2021nyo}. These corrections were obtained
within the threshold resummation formalism, expanding the all-order resummed results up to NNLO accuracy in the strong coupling.
They are approximate in the sense that they contain all the dominant contributions associated
with the emission of soft gluons near threshold (and even some sub-dominant contributions),
but do not yet constitute the full NNLO results. 
Nevertheless, they should be readily suited for an initial phenomenological NNLO analysis of SIDIS data in terms of FFs,
provided that one carefully considers potential limitations in their applicability.

There are several factors that motivate us to perform this study. First, the precision of current LHC experiments and much 
anticipated measurements with identified hadrons at the future Electron-Ion Collider (EIC) \cite{AbdulKhalek:2021gbh}
will make the development of a framework for a full global analysis of FFs at NNLO accuracy,
and encompassing all available probes, mandatory. Although this goal is still rather far off (for example,
NNLO corrections to hadron production in $pp$ scattering are not yet available; see, however, \cite{Hinderer:2018nkb,Gehrmann:2022cih}
for recent progress), we believe that adding SIDIS to the NNLO framework marks an important step in this direction,
demonstrating that global NNLO analyses of FFs are possible in principle. Second, already at this stage
one can examine the question whether the very modest NNLO effects seen in the SIA-only analysis
also recur when adding SIDIS data to the analysis. This addresses the rather important question 
of perturbative stability of the global extraction of FFs. Finally, based on our first exploratory analysis  
we will also be able to delineate the kinematic regions where the NNLO corrections to SIDIS matter most.
This may also shed light on the question in what regions the approximate NNLO terms derived in 
\cite{Abele:2021nyo} are sufficiently accurate, or where a full NNLO calculation for SIDIS may be required. 

{\it Scope and Setup of the Analysis.---} 
Pions are the most copiously produced hadrons and, hence, the corresponding data sets are the most precise ones at hand.
The {\sc Compass} \cite{COMPASS:2016xvm}
and {\sc Hermes} \cite{HERMES:2012uyd} SIDIS results are so far presented in terms of multiplicities, i.e., they are 
normalized to the fully inclusive deep-inelastic cross section. 
While all {\sc Compass} data were taken in scattering off deuterium, {\sc Hermes} has presented results both for proton and deuterium targets.
The two fixed-target experiments have been performed at rather moderate center-of-mass system (c.m.s.) energies $\sqrt{s}$. 
As a consequence, they cover only a fairly limited kinematical range in the relevant variables for deep-inelastic scattering
(DIS), the momentum transfer squared $Q^2$ from the leptonic to the hadronic system, which also sets the relevant
energy scale in DIS and SIDIS, and the Bjorken variable $x$. The latter is related to the momentum fraction of the 
target nucleon that is carried by the initial parton entering the hard scattering.
Semi-inclusive DIS is characterized by an additional variable $z$, defined as the fraction of the virtual photon's energy 
carried by the observed hadron in the target rest frame. Since the bulk of the available SIDIS data sits at very moderate $Q^2$, we 
might expect NNLO corrections to be more pronounced than for SIA where most data were taken on the $Z$-boson resonance.
There are also plenty of SIDIS data at moderate-to-large values of $x$ and/or $z$ where the threshold approximation
for the NNLO cross sections should work very well. 

For consistency, we disregard in what follows the wealth of $pp$ data with identified pions \cite{ref:ppdata} from the 
{\sc Phenix}, {\sc Star}, and {\sc Alice} experiments at the BNL-RHIC and the CERN-LHC, respectively, 
since the relevant NNLO corrections are not yet known. 
In existing NLO fits, these data serve as the main constraint for the gluon-to-pion FF which is otherwise
largely unconstrained. This implies that our obtained set of NNLO pion FFs solely from SIA and SIDIS probes
is not meant to replace the results of existing NLO global analyses, although it may still be useful
for calculations requiring FFs at NNLO.

In the technical analysis, 
we closely follow the general framework outlined and used in previous global analyses of FFs
by the DSS group \cite{deFlorian:2007aj,deFlorian:2017lwf,Borsa:2021ran}. Specifically, the
fragmentation of a parton of flavor $i$ into a positively charged pion is parametrized 
at an initial scale of $Q_0=1\,\mathrm{GeV}$ as
\begin{equation}
\label{eq:ff-input}
D_i^{\pi^+}\!(z,Q_0) =
N_i\,\frac{\sum_{j=1}^3 \,\gamma_{ij}\,z^{\alpha_{ij}}(1-z)^{\beta_{ij}}}{\sum_{j=1}^3 \,\gamma_{ij}\,B(2+\alpha_{ij},1+\beta_{ij})}\,.
\end{equation}
The Euler Beta function $B(a,b)$ normalizes the parameter $N_i$ for each flavor $i$ to its contribution to the 
energy-momentum sum rule. 
As usual, the FFs for negatively charged pions are obtained by 
applying charge conjugation symmetry, $D^{\pi^+}_q=D^{\pi^-}_{\overline{q}}$. 

Since the data cannot constrain all the free parameters in Eq.~(\ref{eq:ff-input}), further assumptions are needed 
\cite{deFlorian:2007aj,deFlorian:2017lwf,Borsa:2021ran}. In particular, with the SIDIS data, the main focus of this paper,
we find a need for a non-vanishing $\gamma_{i3}$ only for $i=u+\bar{u}\equiv u_{\mathrm{tot}}$ and $i=d+\bar{d}\equiv d_{\mathrm{tot}}$.
The $u_{\mathrm{tot}}$ and $d_{\mathrm{tot}}$ FFs are identical up to some possible small degree of charge symmetry breaking,
which we permit by introducing a $z$-independent factor, i.e., 
$D^{\pi^+}_{d_{\mathrm{tot}}} = N_{d_{\mathrm{tot}}}\, D^{\pi^+}_{u_{\mathrm{tot}}}$.
The unfavored light sea quark FF is assumed to obey isospin symmetry, $D^{\pi^+}_{\overline{u}}=D^{\pi^-}_d$, and for the
(anti)strange quark FFs we set $D^{\pi^+}_{s} = D^{\pi^+}_{\bar{s}} =N_s\,z^{\alpha_s}\,D^{\pi^+}_{\overline{u}}$.
As the gluon FF $D^{\pi^+}_g$ is only weakly constrained (see above),
we stick to a simple functional form with $\gamma_{g2}=\gamma_{g3}=0$.

For the present exploratory study, we will refrain from obtaining and quoting error estimates for the FFs but will focus 
instead on a comparison to a reference NLO fit based on the same assumptions, data sets, and functional form (\ref{eq:ff-input})
as outlined above to quantify the impact of the (approximate) NNLO corrections.
Likewise, we do not investigate the reduction of the factorization and renormalization scale uncertainties at NNLO accuracy 
and set all scales equal to the relevant experimental value of $Q$ throughout. 
The scale dependence was studied in Ref.~\cite{Abele:2021nyo}, see Fig.\ 3, and found to be noticeably smaller at NNLO than at NLO 
for typical {\sc Compass} kinematics and $z\gtrsim 0.3$. 

\begin{table*}[t]
\begin{tabular}{ | l || c c c | c c c | c c c | c c c | }
 \hline
 Experiment & \multicolumn{3}{c|}{$Q^2 \geq1.5\,\mathrm{GeV}^2$}& \multicolumn{3}{c|}{$Q^2 \geq2.0\,\mathrm{GeV}^2$}& \multicolumn{3}{c|}{$Q^2 \geq2.3\,\mathrm{GeV}^2$} & \multicolumn{3}{c|}{$Q^2 \geq3.0\,\mathrm{GeV}^2$} \\
 
 \cline{2-4}  \cline{5-7}  \cline{8-10} \cline{11-13}
 
 & \#data & NLO & NNLO & \#data & NLO & NNLO & \#data & NLO & NNLO & \#data & NLO & NNLO \\
 \hline
 
  SIA & 288 & 1.05 & 0.96 & 288 & 0.91 & 0.87 & 288 & 0.90 & 0.91 & 288 & 0.93 & 0.86 \\
  {\sc Compass} & 510 & 0.98 & 1.14 & 456 & 0.91 & 1.04 & 446 & 0.91 & 0.92 & 376 & 0.94 & 0.93 \\
  {\sc Hermes} & 224 & 2.24 & 2.27 & 160 & 2.40 & 2.08 & 128 & 2.71 & 2.35 & 96 & 2.75 & 2.26 \\
 \hline
 \hline 
  \textbf{TOTAL} & 1022 & 1.27 & 1.33 & 904 & 1.17 & 1.17 & 862 & 1.17 & 1.13 & 760 & 1.16 & 1.07 \\
 \hline

\end{tabular}
\caption{\label{tab:exppiontab_Q2cut} Partial and total $\chi^2$ values per data point 
for the sets included in our NLO and NNLO global fits for different lower cuts on $Q^2$.}
\end{table*}

The optimum values for the remaining 26 free parameters $N_i$, $\alpha_{ij}$, $\beta_{ij}$, and $\gamma_{ij}$ in Eq.~(\ref{eq:ff-input}) are
determined using a standard $\chi^2$ minimization procedure. All numerical calculations are efficiently performed in Mellin $N$-moment space,
where also the approximate NNLO corrections of \cite{Abele:2021nyo} were derived. 
The scale evolution of the FFs up to NNLO and the computation of the SIA cross sections utilizes the codes developed in 
Refs.~\cite{Anderle:2015lqa,Vogt:2004ns}. We note that for the NNLO fit we also compute the inclusive DIS cross section
in the denominator of the SIDIS multiplicity at NNLO, using the numerical implementation in \cite{Bertone:2013vaa}. 
Unless stated otherwise, we adopt the recent NNPDF4.0 \cite{Ball:2021leu} set of PDFs
in all our calculations, along with its strong coupling $\alpha_s$, 
and consider the quoted uncertainties in quadrature in the value for $\chi^2$. 

In the case of the approximate NNLO corrections for SIDIS \cite{Abele:2021nyo}, which are given in terms of double Mellin $N,M$ moments,
the relevant structure function $\mathcal{F}(x,z,Q^2)$ involves a double Mellin inverse transform. 
Schematically, it reads
\begin{equation}
    \mathcal{F}(x,z,Q^2)=\int_{\mathcal{C}_N}\frac{dN}{2\pi i}\,x^{-N}\int_{\mathcal{C}_M}\frac{dM}{2\pi i}\,z^{-M}\, \mathcal{F}(N,M,Q^2),
\end{equation}
where $\mathcal{C}_N$ and $\mathcal{C}_M$ denote suitable integration contours in the complex plane.
To facilitate the computational burden, the $N$ integration, related to the PDF dependence of the SIDIS cross section, can
be carried out once prior to the actual fit and stored in grids \cite{Stratmann:2001pb}. For one choice of factorization scale 
and set of PDFs and all 734 SIDIS data points (as in \cite{deFlorian:2007aj,deFlorian:2017lwf,Borsa:2021ran} 
we require $Q^2\geq1.5\,\mathrm{GeV}^2$), this preparatory step takes a few hours 
on a standard multi-core PC. A typical $\chi^2$ minimization at NNLO accuracy to determine the optimum set of FFs is then performed 
in about 20 minutes on a single CPU core. Clearly, our efficient Mellin technique will allow for a full-fletched NNLO global analysis
of FFs in the future including also $pp$ scattering at NNLO accuracy and a detailed estimate of uncertainties based on hundreds of replicas.

{\it Results and Discussion.---} In Table~\ref{tab:exppiontab_Q2cut} we present the results of a series of global fits 
at NLO and NNLO accuracy to the combined set of SIA and SIDIS data with gradually increasing 
cuts on $Q^2$ from its minimum value of $1.5\,\mathrm{GeV^2}$. 
We observe that for $Q^2\geq 1.5$ GeV$^2$ the description of the SIDIS data deteriorates by including the approximate
NNLO corrections. This is particularly noticeable in case of {\sc Compass}, whose $\chi^2$ per data point increases from 0.98 at NLO to 1.14 at NNLO.
However, discarding bins with low values of $Q^2$ systematically leads to a better global fit at NNLO, surpassing the quality of the NLO result 
in terms of the overall $\chi^2$ already for $Q^2\geq 2$ GeV$^2$. Once we demand $Q^2\ge 3\,\mathrm{GeV}^2$,
the NNLO fit shows not only a much improved total $\chi^2$ value, but also a better quality of the description of both 
SIDIS data sets. Interestingly, the NNLO fit shows significant improvements also in case of the SIA data which all sit above the $Q^2$ cuts
we have implemented, in contrast to what was found previously in fits based solely on SIA data \cite{Anderle:2015lqa}. 
This is to be attributed to the additional ``pull'' by the SIDIS data.

Any lower cut in $Q^2$ is kinematically correlated with the lowest value of $x$ accessible in the two SIDIS experiments.
For instance, demanding $Q^2\ge 3\,\mathrm{GeV}^2$ removes all {\sc Compass} data in the range $0.004\le x\le 0.02$, and about half
of the data with $0.02\le x\le 0.06$ from the analysis. Due to the lower c.m.s.\ energy of {\sc Hermes}, its number of 
data points is reduced even more strongly when increasing the lower $Q^2$ cut. 
Previous global analyses \cite{deFlorian:2007aj,deFlorian:2017lwf,Borsa:2021ran} suggest that there are
tensions between the {\sc Hermes} and {\sc Compass} SIDIS data sets. It appears that this tension is
somewhat mitigated by inclusion of the NNLO terms and also by increasing the lower cut on $Q^2$, although 
it still persists at some level, as seen from the relatively larges values of $\chi^2$ relative to the number of data points
for {\sc Hermes}.

\begin{figure}[h]
\begin{center}
\epsfig{figure= 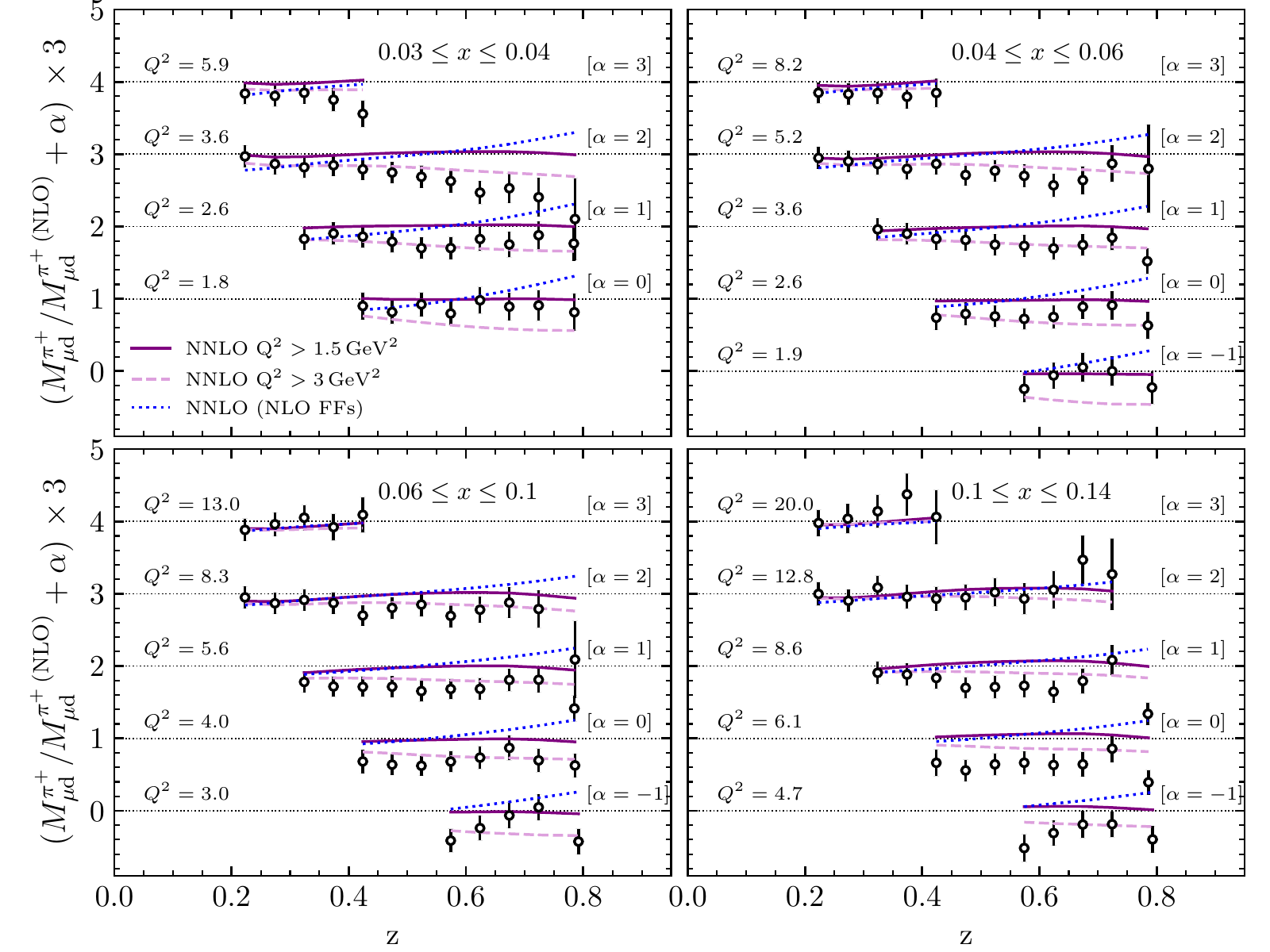 ,width=0.48\textwidth}
\end{center}
\vspace*{-0.4cm}
\caption{Comparison of our NLO and NNLO fits with $Q^2\ge 1.5\,\mathrm{GeV}^2$
to the {\sc Compass} $\pi^+$ multiplicities for some representative bins of $x$,
normalized to the NLO results. The dashed lines show the 
change of the NNLO fit when the $Q^2\ge 3\,\mathrm{GeV}^2$ cut is implemented. 
For a ``test calculation'' represented by the 
dotted lines we have used the FFs from the NLO fit in the NNLO calculation 
(see text). For better clarity, all results are scaled and shifted by constant factors.}
\label{fig:compass_ratios}
%
\begin{center}
\epsfig{figure= 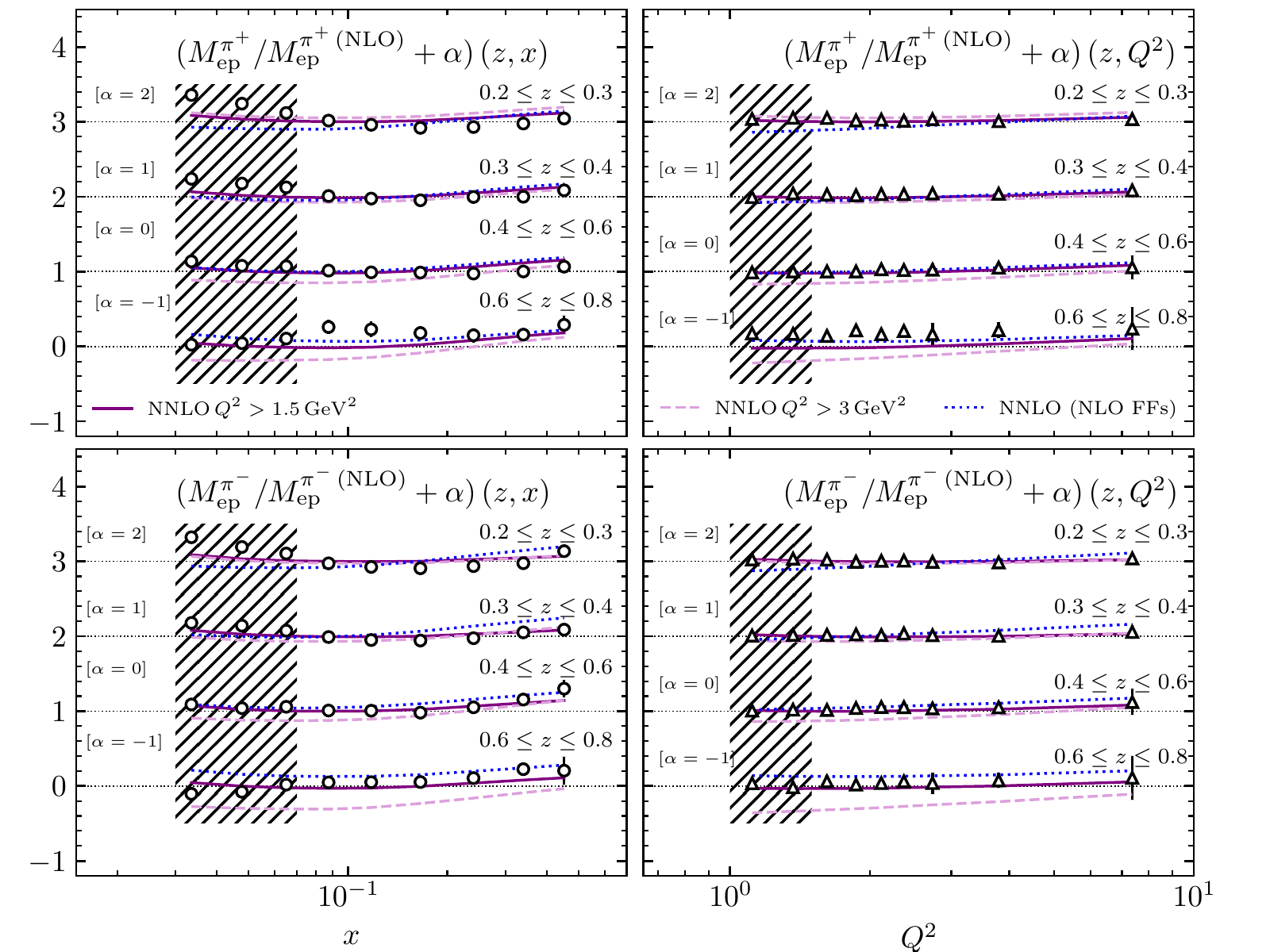 ,width=0.48\textwidth}
\end{center}
\vspace*{-0.3cm}
\caption{Same as in Fig.~\ref{fig:compass_ratios} but now for the {\sc Hermes} $\pi^\pm$ SIDIS data 
on a proton target, for two different projections of the data. 
The shaded areas indicate the regions excluded from the fit by the $Q^2\ge 1.5\,\mathrm{GeV}^2$ cut. 
\label{fig:hermes_ratios}}
\end{figure}

Figures~\ref{fig:compass_ratios} and \ref{fig:hermes_ratios} compare our NLO and NNLO results to the {\sc Compass} 
and {\sc Hermes} SIDIS multiplicities in a few representative bins in $x$. We normalize
all results to the baseline NLO fit with $Q^2\ge 1.5\,\mathrm{GeV}^2$. Using this cut also at NNLO
tends to give a result that slightly overestimates the SIDIS multiplicities, especially at lower values of $x$ or $Q^2$,
which is the reason for the larger $\chi^2$ value seen for this cut in Tab. \ref{tab:exppiontab_Q2cut}.
This changes once we increase the cut to $Q^2\ge 3.0\,\mathrm{GeV}^2$, also shown for NNLO 
in Figs.~\ref{fig:compass_ratios} and \ref{fig:hermes_ratios}, where a clearly improved 
description of the multiplicities is observed. Note that we show the NNLO results for this cut
even for the $Q^2$ values below $3.0\,\mathrm{GeV}^2$, so that the deterioration in the region 
not included in that fit can be seen. 

It is very encouraging that our NNLO analysis based on the approximate NNLO
corrections for SIDIS shows an overall improvement in $\chi^2$ relative to NLO once we go
beyond $Q^2=2\,\mathrm{GeV}^2$. It is an interesting question, however, why the situation
is opposite when the lower cut $Q^2\ge 1.5\,\mathrm{GeV}^2$ is used. 
We first note that the lack of improvement at NNLO when low $Q^2$ are admitted,
as well as the progressive improvements with stricter cuts on $Q^2$,
are not due to the tensions between the different sets of SIDIS data mentioned above. In fact, 
similar results as in Tab.~\ref{tab:exppiontab_Q2cut}
are also obtained in fits using exclusively SIA and {\sc Compass} data. 

Two other possible explanations come to mind.
Clearly, values of $Q^2$ below 3 GeV$^2$ or even 2 GeV$^2$ raise concerns about the 
applicability of a leading-power factorized framework that describes the cross section in terms
of just PDFs, FFs, and perturbative hard-scattering cross sections. It is quite possible that
the trends seen at the lowest $Q^2$ values indicate the onset and perhaps even dominance of power corrections,
very little about which is known theoretically for SIDIS. In a similar vein, at low $Q^2$ one may question
the dominance of the current fragmentation regime~\cite{Boglione:2022gpv}. Even at a practical level the region 
$Q^2< 2\,\mathrm{GeV}^2$ is tedious in our analysis: All modern sets of PDFs incorporate 
a fairly large $Q^2$ cut on the fixed-target DIS data entering their global analysis. 
When computing the SIDIS multiplicities we therefore need to resort to extrapolations of the 
PDF sets outside the region of their applicability for a considerable amount
of the available data, in particular, for {\sc Hermes} kinematics \cite{Borsa:2017vwy}. 
The cut $Q^2\ge 3.0\,\mathrm{GeV}^2$ helps to mitigate this otherwise unavoidable
--- and hard to quantify -- ambiguity stemming from the choice of PDFs. We have verified, however, that replacing
the NNPDF4.0 set \cite{Ball:2021leu}, adopted throughout our work, by the latest MSHT set of PDFs \cite{Bailey:2020ooq}
does not significantly change our results and conclusions. 

The other possibility is that the approximate NNLO corrections for SIDIS might miss some significant 
contributions not associated with the threshold regime. Since threshold resummation only addresses
the region of large $x$ and $z$ such non-threshold contributions, if sizable, could indeed make
the approximate NNLO description unreliable for low values of $x$ and hence $Q^2$. To see
whether there are indications of such a behavior, we have redone the NNLO calculation with
$Q^2\ge 1.5\,\mathrm{GeV}^2$, but using the FFs obtained in the {\it NLO} fit. The corresponding
results are also shown in Figs.~\ref{fig:compass_ratios} and \ref{fig:hermes_ratios}. 
Here the idea is that the ratio to the NLO 
multiplicity will help identify any regions where sizable NNLO corrections arise just from the partonic 
hard-scattering cross sections. As one can see, for large values of $z$ there are significant enhancements 
at NNLO. The threshold logarithms are precisely expected to generate such enhancements.
However, we also observe downward NNLO corrections at lower values of $z$, and here
especially at lower $Q^2$ and hence $x$. Clearly, based on this feature alone we cannot
judge whether this reduction of the cross section at NNLO is an artifact of the 
near-threshold approximation. This would only become possible with a future full NNLO calculation for SIDIS.
For now we just issue a word of caution concerning this point. 

\begin{figure}
\begin{center}
\epsfig{figure= 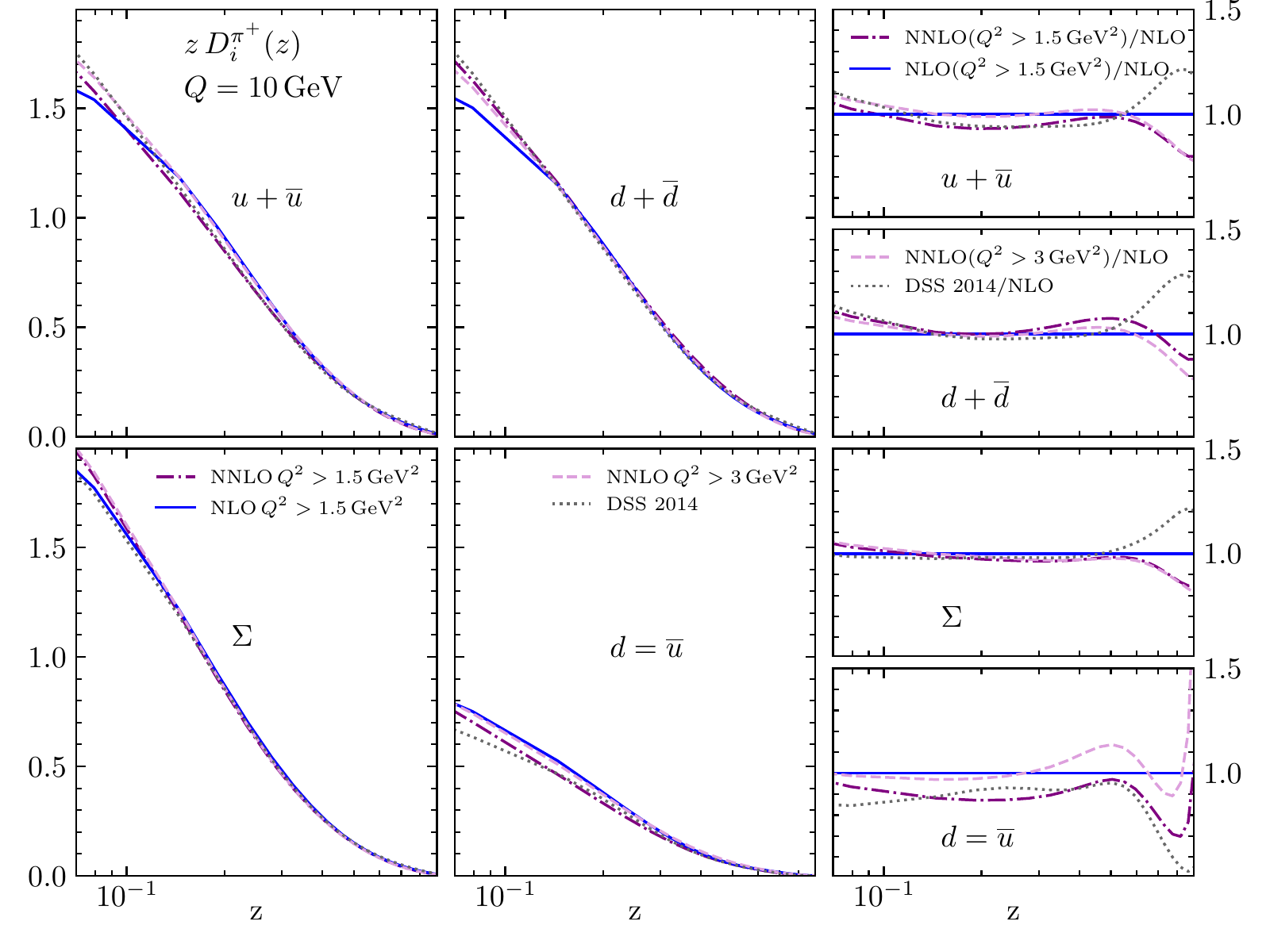 ,width=0.48\textwidth}
\end{center}
\vspace*{-0.3cm}
\caption{Our NLO and NNLO $\pi^+$ FFs for $u_{\mathrm{tot}}$, $d_{\mathrm{tot}}$, $\bar{u}$, and the flavor singlet combination $\Sigma$
at $Q^2=100\,\mathrm{GeV}^2$, compared to those from the DSS analysis of Ref.~\cite{deFlorian:2014xna}. 
The rightmost panels give the ratios to our NLO result. 
\label{fig:dists_100}}
\end{figure}

We finally consider the FFs that our fits produce. Figure~\ref{fig:dists_100} shows the functions for $\pi^+$ 
production, evolved to $Q^2=100\,\mathrm{GeV}^2$, for $u_{\mathrm{tot}}$, $d_{\mathrm{tot}}$, $d=\bar{u}$, and the flavor singlet combination $\Sigma$.
For NNLO, we show results both for $Q^2\ge 1.5\,\mathrm{GeV}^2$ and $Q^2\ge 3\,\mathrm{GeV}^2$ in the fit,
while at NLO we only consider the case $Q^2\ge 1.5\,\mathrm{GeV}^2$. 
All distributions shown are well constrained by our new global fit to SIA and SIDIS data at NNLO accuracy,
while the gluon and strange quark FFs remain largely undetermined. 
The rightmost panels of Fig.~\ref{fig:dists_100}, which present the ratios of the NNLO quark FFs over the NLO ones,
best illustrate the distinct, $z$-dependent pattern of modifications needed when switching 
from NLO to NNLO accuracy. The most significant feature at NNLO is the suppression of the FFs for $z\gtrsim 0.6$,
needed to counteract the enhancements in the partonic cross sections at large $z$ due to the threshold logarithms. 
We note that the cut $Q^2\ge 3.0\,\mathrm{GeV}^2$ mainly leads to an enhancement of $u_{\mathrm{tot}}$ below $z\sim 0.6$
and an additional largely $z$ independent  increase of the unfavored sea quark FFs.

{\it Conclusions.---} 
We have presented the first next-to-next-to-leading order global analysis of parton-to-pion fragmentation functions, 
based on the existing data on single-inclusive electron-proton annihilation and semi-inclusive deep-inelastic scattering. 
The cross sections for the latter were obtained using approximate NNLO corrections previously derived 
within the threshold resummation formalism. Our study is motivated primarily by the desire to improve the theoretical framework 
for the precision analysis
of hadron production data. This task appears particularly important in view of the future EIC that is now on the horizon. 

The FFs we find at NNLO are overall close to the NLO ones, indicating good perturbative stability of the processes used
for the extraction of PDFs. This is especially true for the flavor singlet and the total up and down quark FFs. As expected,
the most noticeable difference at NNLO is the depletion of the FFs in the high-$z$ region, which compensates the increase 
in the partonic cross sections resulting from the NNLO threshold corrections for SIDIS.

We have found that the NNLO corrections improve the overall quality of the fit to the data, 
but do so only when a lower cut of at least $Q^2\geq 2$ GeV$^2$ 
is implemented. With a yet more stringent cut $Q^2\geq 3$ GeV$^2$ the NNLO fit becomes markedly better than the NLO one. This may indicate
that the low-$Q^2$ regime in SIDIS is not well suited for an analysis in terms of factorized cross sections. 
The extrapolation of the PDFs beyond the region where they are constrained by data might be an additional source of inconsistencies. 
That said, also the study of NNLO
corrections not directly associated with the threshold regime will deserve further attention in the future. Ultimately a full NNLO calculation
for SIDIS will be required to describe future EIC data which will likely be taken in kinematic regions quite far away from the threshold regime. 
In any case, we believe that our study is an important step on the way to future fully global NNLO analyses of FFs that include also 
data from hadron production in $pp$ collisions. 

%
{\it Acknowledgements.---} 
I.B.\ wishes to thank 
the University of T\"{u}bingen for hospitality during the completion of the work.
This work was supported in part by CONICET, ANPCyT, UBACyT, and
by Deutsche Forschungsgemeinschaft (DFG) 
through the Research Unit FOR 2926 (Project No. 40824754). 
%
%

\end{document}